\documentclass[prl,pdftex,aps,10pt]{revtex4-1}
\usepackage{graphicx}
\usepackage{amsmath}
\usepackage{graphicx}

\begin{document}
\title{General Charge Balance Functions, A Tool for Studying the Chemical Evolution of the Quark-Gluon Plasma}
\author{Scott Pratt}
\affiliation{Department of Physics and Astronomy and National Superconducting Cyclotron Laboratory,
Michigan State University\\
East Lansing, Michigan 48824}
\date{\today}
\begin{abstract}
In the canonical picture of the evolution of the quark-gluon plasma during a high-energy heavy-ion collision, quarks are produced in two waves. The first is during the first fm/c of the collision, when gluons thermalize into the quark-gluon plasma (QGP). After a roughly isentropic expansion that roughly conserves the number of quarks, a second wave ensues at hadronization, 5-10 fm/c into the collision. Since entropy conservation requires the number of quasi-particles to stay roughly equal, and since each hadron contains at least two quarks, the majority of quark production occurs at this later time. For each quark produced in a heavy-ion collision, an anti-quark of the same flavor is created at the same point in space-time. Charge balance functions identify, on a statistical basis, the location of balancing charges for a given hadron, and given the picture above one expects the distribution in relative rapidity of balancing charges to be characterized by two scales. After first demonstrating how charge balance functions can be defined using any pair of hadronic states, it will be shown how one can identify and study both processes of quark production. Balance function observables will also be shown to be sensitive to the charge-charge correlation function in the QGP. By considering balance functions of several hadronic species, and by performing illustrative calculations, this class of measurement appears to hold the prospect of providing the field's most quantitative insight into the chemical evolution of the QGP.
\end{abstract}

\maketitle

\section{Introduction and Theory}
In a central heavy ion collision at RHIC (Relativistic Heavy Ion Collider) or at the LHC (Large Hadron Collider), several thousand hadrons are created from the initial collision of a few hundred incoming nucleons. In the central unit of rapidity, aside from a few dozen extra baryons, the roughly one thousand hadrons are created and evolve from quark-antiquark creation processes. For every up, down or strange quark observed in the final state, one usually finds one extra anti-up, anti-down or anti-strange antiquark within roughly a unit of rapidity. For quark-antiquark pairs created early in the collision, the balancing pairs might be pulled apart by the initial tunneling process by a fraction of a fm in distance, and are then pulled further apart by collective longitudinal flow and diffusion. If the quarks are pulled a half fm apart at a time 1.0 fm/$c$, collective flow would pull them apart by 7.5 fm by the time breakup occurs ($\sim 15$ fm/$c$), and diffusion would spread them apart even further. In the canonical view of the quark-gluon plasma (QGP), a first wave of quark production occurs when the quark-gluon plasma is created during the first fm/$c$. The number is then roughly conserved in an semi-isentropic expansion until hadronization, when a second wave of production ensues. Due to entropy conservation, a thousand partons would be expected to convert to roughly a thousand hadrons, and since each hadron has multiple quarks, and since the gluonic entropy has no quarks, the number of quarks should more than double during hadronization. If hadronization were to occur later in the process, perhaps at 5-10 fm/$c$, the balancing quark anti-quark pairs created at hadronization would be unlikely to separate by more than a few fm before breakup.

Charge balance functions were proposed as a means for identifying and quantifying the separation of balancing charges \cite{Bass:2000az}. They represent the conditional probability of observing a balancing charge in bin $p_1$ given the observation of a charge in bin $p_2$, and are defined by:
\begin{equation}
B_{+-}(p_1|p_2)\equiv\frac{\left\langle[n_+(p_1)-n_-(p_1)][n_{-}(p_2)-n_+(p_2)]\right\rangle}
{\left\langle n_+(p_2)+n_-(p_2)\right\rangle},
\end{equation}
where $\langle n_{+/-}(p_1)\rangle$ is the probability density for observing a positive/negative particle in bin $p_1$, and $\langle n_+(p_1)n_-(p_2)\rangle$ is the probability density for observing a positive particle in bin $p_1$ and a negative particle in $p_2$. If the number of positives and negatives are equal, and if the detector was perfectly efficient for all $p_1$, integrating the balance function over all $p_1$ would give unity. The label $p_i$ can refer to any measure of momentum, including rapidity or pseudo-rapidity. The observable can be modified to more appropriately treat the case where the net charge significantly differs from zero \cite{Pratt:2003gh}. In short, the balance function is simply the application of a like-sign subtraction with the purpose of statistically isolating the opposite balancing charge.

More generally, balance functions can be analyzed for any two set of hadrons and antihadrons,
\begin{equation}
B_{\alpha\beta}(p_1|p_2)\equiv\frac{\left\langle[n_{\alpha}(p_1)-n_{\bar{\alpha}}(p_1)][n_{\beta}(p_2)-n_{\bar{\beta}}(p_2)]\right\rangle}
{\left\langle n_{\beta}(p_2)+n_{\bar{\beta}}(p_2)\right\rangle},
\end{equation}
For instance, one could consider balance functions where $\alpha$ were protons and $\beta$ were negative kaons. The antiparticles are noted by $\bar{\alpha}$ and $\bar{\beta}$. For this study we will confine ourselves to the situation where the net charges are zero, which is certainly a good assumption at LHC energies. For the case where the net charges are not equal, one might wish to follow the example in \cite{Pratt:2003gh} and define the denominator using the lesser of the two charges, $n_{\bar{\alpha}}$ or $n_\alpha$, followed by a mixed-event subtraction.

Balance functions can be analyzed in six dimensions as a function of $p_1$ and $p_2$, though statistics make that prospect unlikely. Instead, the condition $p_2$ is usually the observation of a particle anywhere in the detector, while $p_1$ refers to the observation of the second particle with relative rapidity $\Delta y$, or relative azimuthal angle $\Delta\phi$, or relative invariant momentum $Q_{\rm inv}$. Balance functions were used at the CERN ISR to study hadronization dynamics in $pp$ and $e^+e^-$ collisions in the 1980s \cite{ISRbalance1,ISRbalance2,ISRbalance3,ISRbalance4,ISRbalance5}, while their use in heavy ion collisions was motivated by the desire to distinguish between early vs. late production of charges \cite{Bass:2000az}. For more central collisions, balance functions have been observed to significantly narrow when binned in relative rapidity \cite{Aggarwal:2010ya,Adams:2003kg,Westfall:2004cq}. This behavior is quantitatively consistent with the idea that a good fraction of the charge is created late in the collision, as expected from delayed hadronization with the existence of a long-lived quark-gluon plasma \cite{Schlichting:2010qia}. Narrowing is also predicted and observed as a function of relative azimuthal angle \cite{Schlichting:2010qia,Bozek:2004dt}, though this paper will focus on the behavior in relative rapidity.

Our first goal is to understand how to calculate balance functions between any two hadronic species $\alpha$ and $\beta$. We especially wish to know what happens if charge production comes in two waves, the first wave being the initial thermalization of the QGP, which is where the quark number rises quickly from zero in the first $\sim 1$ fm/$c$, and the second wave at hadronization, which may be in the 5-10 fm/$c$ window. This goal is complicated by the fact that hadrons carry three charges (one for each light flavor of quark or equivalently strangeness, electric charge and baryon number), which makes the problem rather entangled. Charge balance functions binned in relative spatial rapidity are characterized by a scale $\sigma_{(\rm QGP)}$ before hadronization, which might well be the greater part of a unit of rapidity. During hadronization a second group of balancing charges is created with relative coordinates characterized by $\sigma_{\rm(had)}\sim 0$. In the next section, it will be shown how one can use local charge conservation overlaid onto an assumption that extra charge within a volume is distributed thermally to derive expressions for both components of the balance functions. From this perspective, all balance functions in relative spatial rapidity will be determined in terms of $\sigma_{\rm(had)}$ and $\sigma_{\rm(QGP)}$, the number of quark species per unit rapidity before hadronization, $dN_{u,d,s}/dy$, and the number of hadrons per unit rapidity, $dN_\alpha/dy$ in the final state. The hadronic yields are experimentally measured and the rapidity density of quarks can be estimated from the total entropy, thus leaving the widths as the least understood quantities.  The balance function in terms of the relative spatial coordinate along the beam axis translates into a balance function in relative rapidity after convoluting with a thermal kinetic distribution and including decays. In the final section, we present illustrative predictions for balance functions in relative rapidity for several species by using a thermal blast-wave model to map between spatial rapidity and momentum space rapidity.

\section{Theory}
Balance functions are related to charge correlations. For the purposes of this derivation the charge densities are considered as a function of the coordinate $\eta$, which describes the longitudinal position in Bjorken Coordinates. 
\begin{eqnarray}
z=\tau\sinh(\eta),&~~&t=\tau\cos(\eta),\\
\nonumber
\tau=\sqrt{t^2-z^2},&~~&\eta=\tanh^{-1}(z/\tau).
\end{eqnarray}
In the absence of longitudinal acceleration a particle moving with the fluid has fixed $\eta$, and aside from diffusion the separation of balancing charges would be fixed in $\Delta\eta$.

Before progressing, we define the charge correlation function,
\begin{eqnarray}
\label{eq:gdef}
g_{ab}(\eta_1,\eta_2)&\equiv&\langle\rho_a(\eta_1)\rho_b(\eta_2)\rangle'\\
\nonumber
&=&\sum_{i\ne j}q_{i,a}\delta(\eta_1-\eta_i)q_{j,b}(\eta_2-\eta_j),
\end{eqnarray}
where the sum over $i\ne j$ covers all particles $i$ and $j$, and the prime notes that the correlation of a charge with itself is subtracted. The indices $a$ and $b$ refer to the specific charge, e.g. net strangeness. For this paper Roman indices will refer to charges, e.g., the net number of up, down or strange quarks, while Greek indices denote specific species, e.g., $\pi^+,p,K^-$. Since a chargeless plasma is being considered, $\langle\rho_a\rangle=0$, and one need not subtract the terms $\langle\rho_a\rangle\langle\rho_b\rangle$. For a hadronic system conserving baryon number, electric charge and strangeness, the index can equivalently sum over the net number of up, down and strange quarks. The charge-charge correlation can be expressed as:
\begin{eqnarray}
\label{eq:gB_singlecharges}
g_{++}(\eta_1,\eta_2)&=&
\left\langle[n_+(\eta_1)-n_-(\eta_1)][n_+(\eta_2)-n_-(\eta_2)]\right\rangle^\prime\\
\nonumber
&=&-B_{+-}(\eta_1|\eta_2)\left\langle n_+(\eta_2)+n_-(\eta_2)\right\rangle.
\end{eqnarray}
Here, the positive and negative subscripts refer to the sets of all positive, or all negative particles. The relation between the balance function, $B_{\alpha\beta}$ and $g_{ab}$ becomes complicated if particles have more varied charges, which is the case for a hadronic system, e.g., the $\Sigma^-$ carries baryon number, electric charge and strangeness. These cases will be discussed in the next few paragraphs.

The reason we switch from balance functions to correlations is that the correlation does not change suddenly at hadronization, except for where $\eta_1=\eta_2$. This follows from local charge conservation. It can be understood by considering the addition of a pair with $\eta_1\simeq\eta_2$ during hadronization. The contribution of this single pair to the sum in Eq. (\ref{eq:gdef}) where either $i$ or $j$ points to any other particle besides those from the created pair vanishes, because one is considering the creation of a pair with equal but opposite charges at the same point. The only contribution comes from the element of the sum where both particles come from the pair, which then shows up at $\eta_1=\eta_2$. Assuming hadronization is sudden, and assuming one understands the charge correlation before hadronization, one would also know $g(\eta_1,\eta_2)$ immediately after hadronization, except for the region $\eta_1\approx\eta_2$. To determine the correlation in the region of small relative coordinate, one can use the sum-rule for charge correlations, which follows from integrating the definition of $g$ in Eq. (\ref{eq:gdef}),
\begin{eqnarray}
\label{eq:gsumrule}
-\int d\eta_1 g_{ab}(\eta_1,\eta_2)&=&-\sum_{i\ne j}q_{j,a}\delta(\eta_2-\eta_j)q_{i,b}\\
&=&\sum_j q_{j,a}q_{j,b}\delta(\eta_2-\eta_j)\\
\nonumber
&=&\chi_{ab}(\eta_2)\equiv\sum_\alpha \langle n_\alpha(\eta_2)\rangle
q_{\alpha,a}q_{\alpha,b}.
\end{eqnarray}
The first step used charge conservation, $\sum_iq_i=0$. The average number of particles of a given species $\alpha$ within $d\eta$ is $\langle n_\alpha(\eta)\rangle d\eta$.  Assuming instantaneous hadronization, in order to satisfy the sum rule of Eq. (\ref{eq:gsumrule}), the charge correlation immediately after hadronization must be:
\begin{eqnarray}
\label{eq:QGPhad}
g_{ab}(\eta_1,\eta_2)&=&g_{ab}^{\rm(QGP)}(\eta_1,\eta_2)+g_{ab}^{\rm(had)}(\eta_1,\eta_2),\\
\nonumber
g^{\rm (had)}_{ab}(\eta_1,\eta_2)&=&-\left[\chi^{(\rm had)}_{ab}(\eta_1)-\chi^{(\rm QGP)}_{ab}(\eta_1)\right]\delta(\eta_1-\eta_2),\\
\nonumber
\chi^{\rm(had)}_{ab}(\eta)&=&\sum_{\alpha\in{\rm had}}q_{\alpha,a}q_{\alpha,b}\langle n_{\alpha}(\eta)\rangle\\
\nonumber
\chi^{\rm(QGP)}_{ab}(\eta)&=&\sum_{\alpha\in{\rm QGP}}q_{\alpha,a}q_{\alpha,b}\langle n_\alpha(\eta)\rangle,
\end{eqnarray}
where $g^{\rm(QGP)}$ describes the correlations both immediately before and immediately after hadronization, but neglects the hadronization component created at $\eta_1=\eta_2$. The sums over $\alpha$ cover the species for each state, i.e., over partonic species for the QGP state and over hadronic species for the hadronic state. The value $\chi_{ab}$, when multiplied by the delta function, represents the charge-charge correlation that would ensue from independent particles, i.e., when the only correlations come from a particle with itself. Here the values $\langle n_\alpha\rangle$ are the densities per unit $\eta$ of the species $\alpha$, so if one measures the final-state yields $\chi_{ab}^{\rm(had)}$ can be considered as known. The values of $\chi_{ab}$ can also be extracted from a one-body treatment such as hydrodynamics. The matrix $\chi^{\rm{(QGP)}}_{ab}$ is diagonal in a QGP if the charges refer to the net number of up, down and strange quarks. In contrast, hadrons have multiple charges and $\chi^{\rm(had)}_{ab}$ has off-diagonal elements. Since hadronization is sudden, but not instantaneous, one would expect to replace the delta function with some function of finite but narrow width, normalized to unity.

Our next goal is to determine the balance function for any hadronic species just after hadronization, given $g_{ab}$ in the QGP phase. Eq. (\ref{eq:QGPhad}) describes how to extract $g_{ab}$ just after hadronization. However, once there are multiple charges spread across a variety of species, it is not easy to understand how the correlation functions, $g_{ab}(\eta_1,\eta_2)$, determine the balance functions, $B_{\alpha\beta}(\eta_1|\eta_2)$. Here, $a$ and $b$ refer to any conserved charges, while $\alpha$ and $\beta$ refer to the charge carried by a specific species, where the particle and anti-particles of each species are denoted by $\alpha$ and $\bar{\alpha}$,
\begin{eqnarray}
\label{eq:Bofg}
B_{\alpha\beta}(\eta_1|\eta_2)&=&\frac{\langle[n_\alpha(\eta_1)-n_{\bar{\alpha}}(\eta_1)][n_{\beta}(\eta_2)-n_{\bar{\beta}}(\eta_2)]\rangle}
{\langle n_{\beta}(\eta_2)\rangle+\langle n_{\bar{\beta}}(\eta_2)\rangle}\\
\nonumber
&=&\frac{g_{\alpha\beta}(\eta_1,\eta_2)}
{\langle n_{\beta}(\eta_2)\rangle+\langle n_{\bar{\beta}}(\eta_2)\rangle}.
\end{eqnarray}
Here, $n_\alpha$ is the density (number per unit $\eta$) of particles of species $\alpha$. Thus, $g_{\alpha\beta}$ is the correlation of the effective charge defined by the number of a specific species minus the number of its antiparticle. With this definition, one can see that
\begin{eqnarray}
g_{\alpha\bar{\beta}}&=&-g_{\alpha,\beta},~~g_{\bar{\alpha}\beta}=-g_{\alpha\beta},~~
g_{\bar{\alpha}\bar{\beta}}=g_{\alpha\beta},\\
\nonumber
B_{\alpha\bar{\beta}}&=&-B_{\alpha\bar{\beta}},~~B_{\bar{\alpha},\beta}=-B_{\alpha\beta},~~
B_{\bar{\alpha}\bar{\beta}}=B_{\alpha\beta}.
\end{eqnarray}
As an example, one can consider the proton-$K^-$ balance function. In this example, the index $\alpha$ would refer to protons and $\beta$ would refer to negative kaons. The corresponding charge correlation function would be
\begin{equation}
g_{pK^-}(\eta_1,\eta_2)=\langle[n_p(\eta_1)-n_{\bar{p}}(\eta_1)][n_{K^-}(\eta_2)-n_{K^+}(\eta_2)]\rangle.
\end{equation}

The suffixes $\alpha$ and $\beta$ can also refer to a subset of species, with $\bar{\alpha}$ and $\bar{\beta}$ referring to the equivalent subset of antiparticles. For instance, $\alpha$ could refer to the set of all positive particles, while $\beta$ could refer to the set of all antiparticles. Switching the indices leads to the relations:
\begin{equation}
g_{\alpha\beta}=g_{\beta\alpha},~~B_{\alpha\beta}n_\beta=B_{\beta\alpha}n_\alpha.
\end{equation}

Determining the balance functions for arbitrary species requires making the jump from $g_{ab}$ to $g_{\alpha\beta}$. There are three conserved charges, which we will consider to be the net numbers of up, down and strange quarks. Although one could have equivalently used baryon number, electric charge and strangeness, the quark numbers are more convenient since one does not expect any off-diagonal elements to $g_{ab}$ in this basis for the QGP. For the species-labeled correlations, $g_{\alpha\beta}$, there are many more possibilities in the hadronic state. Even for the final state, one might wish to consider charged pions, charged kaons, protons or lambdas. Neutral kaons must also be taken into account for absorbing strangeness, but because they oscillate into $K_s$ and $K_l$, cannot be easily used for balance functions. Since $g_{\alpha\beta}$ has more elements than $g_{ab}$, additional assumptions are required if $g_{\alpha\beta}$ is to be determined from $g_{ab}$. 

Observing a hadronic species $\alpha$ at position $\eta_1$ infers one has observed the three charges $q_{\alpha,a}$, which is the number of up, down and strange quarks in the resonance $\alpha$. The correlation $g_{ab}(\eta_1,\eta_2)$ should then provide the probability of finding the balancing charges at position $\eta_2$. In order to deterine $g_{\alpha\beta}$ one then needs a model to determine how an extra charge $q_b$ at position $\eta_2$ influences the probability of finding a hadronic species $\beta$ at the same position.

By assuming that the local distribution of hadrons is determined by a thermal distribution constrained by the local charge density, one can determine $g_{\alpha\beta}$ from $g_{ab}$. To show this, we express the two particle correlation as being determined by a grand canonical ensemble with Lagrange multipliers applied to constrain reproduction of the average two-particle correlation function, i.e.,
\begin{eqnarray}
\label{eq:AB}
\langle AB\rangle&=&\frac{1}{Z}{\rm Tr}\left\{ABe^{-\int d\eta H_0/T(\eta)}\exp\left[\int d\eta_1d\eta_2~
\sum_{ab}\rho_a(\eta_1)\mu_{ab}(\eta_1,\eta_2)\rho_b(\eta_2)\right]\right\},\\
\nonumber
Z&=&{\rm Tr}\left\{e^{-\int d\eta H_0/T(\eta)}\exp\left[\int d\eta_1d\eta_2~\sum_{ab}\rho_a(\eta_1)\mu_{ab}(\eta_1,\eta_2)\rho_b(\eta_2)\right]\right\}.
\end{eqnarray}
Here,  $H_0$ is the Hamiltonian or relevant free energy density, $T$ is the temperature, and $\mu_{a,b}(\eta_1,\eta_2)$ plays the role of a Lagrange multiplier chosen to enforce that $g_{ab}(\eta_1,\eta_2)$ is reproduced. The strategy will be first to find $\mu_{ab}$ in terms of $g_{ab}$, then to use $\mu_{ab}$ to determine $g_{\alpha\beta}$. The correlation function $g_{ab}(\eta_1,\eta_2)$ is found by replacing the operators $A$ and $B$ above with
\begin{equation}
A=\rho_a(\eta_1)=\sum_\alpha n_\alpha(\eta_1) q_{\alpha,a}, ~~~
B=\rho_b(\eta_2)=\sum_\beta n_\beta(\eta_2) q_{\beta,b},
\end{equation}
where $\alpha$ and $\beta$ are summed over all hadronic species. By assuming that the weighting is proportional to an exponential of the constraint (fixing $g_{ab}$), this is essentially a thermal ansatz. 

Since the correlation would be zero if not for $\mu$, we can expand the expression for small $\mu$ and find:
\begin{eqnarray}
g_{ab}(\eta_1,\eta_2)&=&\sum_{\alpha\beta}\langle n_\alpha(\eta_1)\rangle q_{\alpha,a}
q_{\beta,b}\langle n_\beta(\eta_2)\rangle
\exp\left\{\sum_{cd}q_{\alpha,c}\mu_{cd}(\eta_1,\eta_2)q_{\beta,d}\right\}\\
\nonumber
&\approx&
\sum_{\alpha\beta cd} \langle n_\alpha(\eta_1)\rangle q_{\alpha,a}q_{\alpha,c}\mu_{cd}(\eta_1,\eta_2)
q_{\beta,d}q_{\beta,b}\langle n_\beta(\eta_2)\rangle,\\
\nonumber
&=&\sum_{cd}\chi_{ac}(\eta_1)\mu_{cd}(\eta_1,\eta_2)\chi_{db}(\eta_2),
\end{eqnarray}
where $\chi$ was defined in Eq. (\ref{eq:gsumrule}). The assumption of small $\mu$ is warranted given that charge-conservation correlations are small (at least for central collisions). Inverting the equation, one can then find $\mu_{ab}$ in terms of $g_{ab}$,
\begin{equation}
\mu_{ab}(\eta_1,\eta_2)=\sum_{cd}\chi^{(-1)}_{ac}(\eta_1)g_{cd}(\eta_1,\eta_2)\chi^{(-1)}_{db}(\eta_2).
\end{equation}

One can now find $g_{\alpha\beta}$ by inserting 
\begin{equation}
A=n_\alpha(\eta_1)-n_{\bar{\alpha}}(\eta_1),~~~B=n_\beta(\eta_2)-n_{\bar{\beta}}(\eta_2),
\end{equation}
into Eq. (\ref{eq:AB}). Here $n_{\bar{\alpha}}$ is the density of the anti-particles to $\alpha$. Again, assuming equal numbers of particles and antiparticles, $\langle n_\alpha\rangle=\langle n_{\bar{\alpha}}\rangle$, and assuming that $\mu_{ab}$ is small,
\begin{eqnarray}
g_{\alpha\beta}(\eta_1,\eta_2)&=&
\left\langle\left[n_\alpha(\eta_1)-n_{\bar\alpha}(\eta_1)\right]\left[n_\beta(\eta_2)-n_{\bar\beta}(\eta_2)\right]\right\rangle\\
\nonumber
&=&\langle n_\alpha(\eta_1)\rangle\langle n_\beta(\eta_2)\rangle
\exp\left\{\sum_{ab}q_{\alpha,a}\mu_{ab}(\eta_1,\eta_2)q_{\beta,b}\right\}
+\langle n_{\bar{\alpha}}(\eta_1)\rangle\langle n_{\bar{\beta}}(\eta_2)\rangle
\exp\left\{\sum_{ab}q_{\alpha,a}\mu_{ab}(\eta_1,\eta_2)q_{\beta,b}\right\}\\
\nonumber
&&-\langle n_{\alpha}(\eta_1)\rangle\langle n_{\bar{\beta}}(\eta_2)\rangle
\exp\left\{-\sum_{ab}q_{\alpha,a}\mu_{ab}(\eta_1,\eta_2)q_{\beta,b}\right\}
-\langle n_{\bar{\alpha}}(\eta_1)\rangle\langle n_{\beta}(\eta_2)\rangle
\exp\left\{-\sum_{ab}q_{\alpha,a}\mu_{ab}(\eta_1,\eta_2)q_{\beta,b}\right\}\\
\nonumber
&\simeq&4\langle n_{\alpha}(\eta_1)\rangle
q_{\alpha,a}\mu_{ab}(\eta_1,\eta_2)q_{\beta,b}\langle n_{\beta}(\eta_2)\rangle.
\end{eqnarray}
From Eq. (\ref{eq:Bofg}), one then finds an expression for the balance function,
\begin{eqnarray}
\label{eq:Bresult}
B_{\alpha\beta}(\eta_1|\eta_2)&=&2\sum_{ab}\langle n_{\alpha}(\eta_1)\rangle
q_{\alpha,a}\mu_{ab}(\eta_1,\eta_2)q_{\beta,b}\\
\nonumber
&=&2\sum_{abcd}\langle n_{\alpha}(\eta_1)\rangle
q_{\alpha,a}\chi^{(-1)}_{ac}(\eta_1)g_{cd}(\eta_1,\eta_2)\chi^{(-1)}_{db}(\eta_2)
q_{\beta,b}.
\end{eqnarray}
One test of this result is to see whether integrating the balance function over all $\eta_1$, summing over all $\alpha$, and weighting with $q_{\alpha,a}$, one should get the net amount of charge $a$ found in other particles due to the condition of having observed a particle of species $\beta$ at position $\eta_2$. Performing these operations from the expression for $B$ in Eq. (\ref{eq:Bresult}),
\begin{eqnarray}
\sum_\alpha \int d\eta_1~q_{\alpha,a}B_{\alpha\beta}(\eta_1|\eta_2)&=&
2\int d\eta_1~\sum_{\alpha bcd}q_{\alpha,a}\langle n_\alpha(\eta_1)\rangle q_{\alpha,b}
\chi^{(-1)}_{bc}(\eta_1)g_{cd}(\eta_1,\eta_2)\chi^{(-1)}_{db}(\eta_2)q_{\beta,b}\\
\nonumber
&=&2\int d\eta_1~\chi_{ab}(\eta_1)\chi_{bc}^{(-1)}(\eta_1)
g_{cd}(\eta_1,\eta_2)\chi^{(-1)}_{db}(\eta_2)q_{\beta,b}\\
\nonumber
&=&\int d\eta_1~
g_{ac}(\eta_1,\eta_2)\chi^{(-1)}_{cb}(\eta_2)q_{\beta,b}\\
\nonumber
&=&-2\sum_{cd}\chi_{ac}(\eta_2)\chi^{(-1)}_{cd}(\eta_2)q_{\beta,d}\\
\nonumber
&=&-2q_{\beta,a}.
\end{eqnarray}
The second-to-last step used the sum rule for integrating $g$ in Eq. (\ref{eq:gsumrule}). The factor of two comes from the fact that the sum over all species, $\alpha$, double-counted the contributions. For instance, the term for which $\alpha=\pi^+$ also includes the contribution from $\pi^-$, and the term for $\alpha=\pi^-$ also includes the contribution from the $\pi^+$.

\section{Calculating Weights for Both Components for All Hadronic Species}

From Eq. (\ref{eq:QGPhad}), one expects two components to the charge correlation $g_{ab}(\eta_1,\eta_2)$. Assuming a boost-invariant system, one can assume a dependence on $\Delta\eta=\eta_1-
\eta_2$, rather than on $\eta_1$ and $\eta_2$ individually. This expectation inspires one to write the balance function for all species $B_{\alpha\beta}(\Delta\eta)$ in terms of two components,
\begin{equation}
\label{eq:weightdef}
B_{\alpha\beta}(\Delta\eta)=w^{\rm (QGP)}_{\alpha\beta}b^{({\rm QGP})}(\Delta\eta)+w^{\rm (had)}_{\alpha\beta}b^{({\rm had})}(\Delta\eta),
\end{equation}
where $b^{({\rm QGP})}$ and $b^{({\rm had})}$ are both normalized so that $\int d\Delta\eta b(\Delta\eta)=1$. 

The weights, $w^{\rm(QGP)}$ and $w^{\rm(had)}$, can be determined from the charge correlations, which in turn depend on the matrices $\chi_{ab}$. From Eq. (\ref{eq:QGPhad}),
\begin{equation}
-g_{ab}(\Delta\eta)=\chi^{\rm(QGP)}_{ab}b^{\rm(QGP)}(\Delta\eta)+\left[\chi^{\rm(had)}_{ab}-\chi^{\rm(QGP)}_{ab}\right]b^{\rm(had)}(\Delta\eta).
\end{equation}
Here, the delta function in Eq. (\ref{eq:QGPhad}) was replaced by a Gaussian of finite width, where the width is determined by the charge diffusion between hadronization and breakup. The correlation before hadronization, $g_{ab}^{\rm(QGP)}$, should be diagonal if quarks are good quasi-particles,
\begin{equation}
\chi_{ab}^{\rm(QGP)}=\langle n_a+n_{\bar{a}}\rangle\delta_{ab},
\end{equation}
where $n_a$ is the density of up, down or strange quarks, and $n_{\bar{a}}$ is the density of the antiquarks. In this formulation there is an explicit assumption that the diffusive widths of the charge correlation before hadronization are independent of flavor. Whereas the form of $\chi^{\rm(QGP)}$ is model dependent, $\chi^{\rm(had)}_{ab}=\langle n_{\alpha}\rangle q_{\alpha,a}q_{\beta,b}$ is determined from final-state yields. After inserting the above expression for $g_{ab}$ into Eq. (\ref{eq:Bresult}), one obtains $B_{\alpha\beta}$, from which one can read off the weights in Eq. (\ref{eq:weightdef}),
\begin{eqnarray}
w_{\alpha\beta}^{\rm(QGP)}&=&-2\sum_{abcd}\langle n_{\alpha}\rangle q_{\alpha,a}\chi_{ab}^{-1{\rm(had)}}
\chi_{bc}^{\rm(QGP)}\chi_{cd}^{-1{\rm(had)}}q_{\beta,d},\\
\nonumber
w_{\alpha\beta}^{\rm(had)}&=&-2\sum_{ab}\langle n_{\alpha}\rangle q_{\alpha,a}\chi_{ab}^{-1{\rm(had)}}q_{\beta,b}
-w_{\alpha\beta}^{\rm(QGP)}.
\end{eqnarray}

The characteristic width of $b^{\rm (QGP)}$ is determined by the charge correlation before hadronization, $g_{ab}^{\rm(QGP)}(\Delta\eta)$, and one might expect it to be of the order $\gtrsim 0.5$. In contrast, $b^{\rm(had)}$ is characterized by a narrow width describing the diffusion of charge after hadronization and might have a width $\simeq 0.1-0.2$. Although the derivations assumed that the species were locally populated according to local thermal equilibrium, the weights are completely determined given the populations for quarks just before hadronization, and the rapidity density for hadronic species $\langle n_\alpha\rangle$.

Whereas the hadronic populations can be taken from experiment (or from a thermal model tuned to experiment), the number density of quarks just before hadronization is dependent on model assumptions. Even if one uses entropy arguments to infer the number of quarks, neglecting the entropy created during hadronization, the number of quarks can depend on how much entropy was carried by gluons. For that reason the ratio of the rapidity density of quarks before hadronization to the rapidity density of final state hadrons was varied. Three ratios were explored: $n_{\rm quarks}/n_{\rm had}=$0.7, 0.85 and 1.0. The hadron density included neutral hadrons, and the decay products of strange baryons and the $K_s$.

Despite the wide coverage and detailed analysis of RHIC data, the uncertainty in the yields of particular species at RHIC can be rather large. Whereas the yields of pions are known to better than the 10\% level, yields of protons and anti-protons are uncertain at the 25\% level. Given these uncertainties, we use yields from a thermal calculation based on a temperature of 165 MeV. The calculation involved generating particles thermally from a hydrodynamic evolution. Particles of all hadronic species were then evolved through a hadronic cascade, whose main purpose was to model the hadronic decays. Since only the yields were sought, the dynamical evolution of the cascade was rather inconsequential. Weak decays were not performed. The remaining species and their yields for central collisions are given in Table \ref{table:hadronyields}. The yields given in the table were then modified by an additional factor $f_B$, which reduced the yields of all baryons by the same factor. Given that the number of anti-baryons is less than the number of baryons at RHIC by a factor of 0.7, and given that only anti-baryons are accompanied by an additional charge, one might expect to a factor of $f_B\approx 0.85$. Comparing the numbers below to proton yields from PHENIX \cite{Adler:2003cb}, one would expect $f_B\approx 0.5$, whereas a value closer to 0.7 might be expected from STAR's yields \cite{Adams:2003xp,Abelev:2008ez}. The cascade code did not include baryon-baryon annihilations, and it is not clear whether a thermal calculation followed by hadronic processes would be more consistent with the STAR or the PHENIX values. Given these uncertainties, calculations were performed for three values, $f_B=0.5, 0.6$ and $0.7$. 

Only a partial list of resonances were applied in the calculation of weights, those that survive to the final state aside from weak decays. For baryons this includes protons, neutrons, $\Sigma^{+/-}$, $\Xi^{-/0}$, $\Omega^-$ and the corresponding antibaryons. The included mesons are $\pi^{+/0/-}$ and $K^{+,-,0}$. The thermal model provides yields, $\langle n\rangle_\alpha$, at midrapidity, and are listed in Table \ref{table:hadronyields}. The ``thermal'' model was a hydrodynamic model followed by a cascade simulation, where hadronization was performed thermally with a temperature of 165 MeV. 
\begin{table}
\begin{tabular}{|c|c|}\hline
hadron species & yields, $n_\alpha$\\ \hline
$p,n,\bar{p},\bar{n}$ & 22.5\\
$\Lambda, \bar{\Lambda}$ & 8.5\\
$\Sigma^+,\Sigma^-,\bar{\Sigma}^-,\bar{\Sigma}^+$ & 3.4\\
$\Xi^-,\Xi^0,\bar{\Xi}^0,\bar{\Xi}^+$ & 1.95\\
$\Omega,\bar{\Omega}$ & 0.35\\
$\pi^+,\pi^0,\pi^-$ & 268\\
$K^+,K^-$ & 54 \\
\hline
\end{tabular}
\caption{\label{table:hadronyields}
Hadronic yields used to calculate weights. Yields were calculated from a hydrodynamic/cascade model with no net baryon number, where the initial hadronic populations were set according to a thermal distribution with a temperature of 165 MeV. An additional factor, $f_B$, was applied to the baryon yields listed here to account for the experimental uncertainties and for greater consistency with experimental observations.}
\end{table}

The resulting weights for the default calculation ($f_B=0.6, ~n_{\rm quarks}/n_{\rm had}=0.85$) are shown in Table \ref{table:default}. The weights for the $\pi^+\pi^-$ balance functions were not surprising. In the default calculation the hadronization process is responsible for nearly two thirds of the final quarks (2 for a meson plus three for a baryon). Pions represent $\sim 80$\% of the final-state particles, and given these facts it was not surprising that the hadronization component of the $\pi\pi$ balance functions integrated to 0.636, while the QGP component integrated to 0.239. The sum did not integrate to unity because observing a charged pion does not ensure that the remainder of the system has one extra pion of the opposite charge due to the possibility of the charge being balanced by other species. If one could measure the balance function in coordinate space, i.e. as a function of $\eta$, there would be a large narrow peak from the hadronization component and a smaller broader structure from the QGP component.

\begin{table}
\begin{tabular}{|r|c|c|c|c|c|c|c|c|c|c|c|} \hline
                  &           $p$ &     $\Lambda$ &    $\Sigma^+$ &    $\Sigma^-$ &       $\Xi^0$ &       $\Xi^-$ &    $\Omega^-$ &       $\pi^+$ &         $K^+$\\ \hline
       $\bar{p}$  &  0.441,-0.066 &  0.485,-0.162 &  0.491,-0.146 &  0.479,-0.178 &  0.535,-0.242 &  0.529,-0.258 &  0.578,-0.338 &  0.006, 0.016 & -0.044, 0.096\\ 
 $\bar{\Lambda}$  &  0.183,-0.061 &  0.242,-0.094 &  0.242,-0.094 &  0.242,-0.094 &  0.302,-0.128 &  0.302,-0.128 &  0.361,-0.161 &  0.000,-0.000 & -0.059, 0.033\\ 
$\bar{\Sigma}^-$  &  0.074,-0.022 &  0.097,-0.038 &  0.099,-0.033 &  0.095,-0.043 &  0.122,-0.049 &  0.120,-0.054 &  0.144,-0.064 &  0.002, 0.005 & -0.023, 0.016\\ 
$\bar{\Sigma}^+$  &  0.072,-0.027 &  0.097,-0.038 &  0.095,-0.043 &  0.099,-0.033 &  0.120,-0.054 &  0.122,-0.049 &  0.144,-0.064 & -0.002,-0.005 & -0.025, 0.011\\ 
   $\bar{\Xi}^0$  &  0.046,-0.021 &  0.069,-0.029 &  0.070,-0.028 &  0.069,-0.031 &  0.093,-0.036 &  0.092,-0.038 &  0.115,-0.045 &  0.001, 0.001 & -0.023, 0.008\\ 
   $\bar{\Xi}^+$  &  0.046,-0.022 &  0.069,-0.029 &  0.069,-0.031 &  0.070,-0.028 &  0.092,-0.038 &  0.093,-0.036 &  0.115,-0.045 & -0.001,-0.001 & -0.023, 0.007\\ 
$\bar{\Omega}^+$  &  0.009,-0.005 &  0.015,-0.007 &  0.015,-0.007 &  0.015,-0.007 &  0.021,-0.008 &  0.021,-0.008 &  0.027,-0.009 & -0.000,-0.000 & -0.006, 0.001\\ 
         $\pi^-$  &  0.119, 0.318 &  0.000,-0.000 &  0.239, 0.636 & -0.239,-0.636 &  0.119, 0.318 & -0.119,-0.318 & -0.000,-0.000 &  0.239, 0.636 &  0.119, 0.318\\ 
           $K^-$  & -0.175, 0.384 & -0.627, 0.352 & -0.603, 0.417 & -0.651, 0.288 & -1.055, 0.385 & -1.079, 0.321 & -1.507, 0.354 &  0.024, 0.064 &  0.452, 0.031\\ 
\hline
\end{tabular}
\caption{\label{table:default}
Default results for the weights $w_{\alpha\beta}^{\rm(QGP)}$, $w_{\alpha\beta}^{\rm(had)}$ resulting from the thermal model described in the text. The weights describe the contribution to the balance functions $B_{\alpha\beta}(\Delta\eta)$ from the correlations driven by the charge correlations just before hadronization and the additional correlation that appears during hadronization and is local, $\Delta\eta\sim 0$.}
\end{table}

The default results for the $K^+K^-$ balance functions were also in line with expectations. Since rather few additional strange quarks are produced during hadronization, the hadronization component turned out to be quite small. Observing the lack of a narrow peak in $K^+K^-$ balance functions would confirm the notion that the QGP was indeed rich in strangeness.

The $p\bar{p}$ balance function came out contrary to expectations expressed in previous papers \cite{Bass:2000az}. Even though protons are composed entirely of up and down quarks, and even though a large fraction of up and down quarks are produced at hadronization, the hadronization component is small, or perhaps negative. This comes from the fact that the strength of the hadronization component, as determined by the sum rule in Eq. (\ref{eq:QGPhad}), depends on the density of observed baryons. Since the observed number of baryons is rather small, the sum rule can be saturated by the number of baryons in the QGP component. If one were to consider baryons alone, the sign of the hadronization component in the baryon-baryon correlation depends on the sign of $\chi_{bb}^{\rm (had)}-\chi_{bb}^{\rm (QGP)}$, where $bb$ would refer to the baryon charge. Since the baryon number of a single quark is 1/3, the sign switches when the number of quarks is more than nine times the number of baryons. 

Another surprising result in Table \ref{table:default} concerns the $pK^-$ balance function. Even though the $K^-$ meson has an anti-up quark, the QGP component is negative. This derives from $\mu_{ab}$ being larger for $us$ than for $uu$. For the range of parameters explored here, the hadronization component of the $pK^-$ balance function was always positive. This makes it easy to recognize the existence of both the QGP and hadronization components, and if such a structure were observed experimentally, it would be difficult to explain without a two-component picture of quark production.

The upper two tables in Table \ref{table:bfudgenq} show the dependence of the weights for variations of the baryon suppression $f_B$, which scales the final baryon yields relative to thermal yields. The values of $f_B$ roughly span the range of uncertainties from the experimental measurement. Whereas the default value of $f_B$ was set to 0.6, Table \ref{table:bfudgenq} shows results for $f_B=0.5$ and $f_B=0.7$. The number of quarks per unit rapidity in the QGP just before hadronization is also uncertain, hence a range of quark numbers is explored. Bracketing the default ratio of quarks before hadronization to final-state hadrons of 0.85, results for $n_{\rm quarks}/n_{\rm had}=0.7$ and $n_{\rm quarks}/n_{\rm had}=1.0$ are shown in the bottom two tables. The $p\bar{p}$ balance function is especially sensitive to both numbers. The hadronization peak is strengthened by raising $f_B$, or by lowering $n_{\rm quarks}/n_{\rm had}$. For lower baryon yields, or higher quark densities, the hadronization peak becomes smaller, and can even become negative. These would lead to a dip in the $p\bar{p}$ balance function at small relative rapidity, which would both provide striking evidence of the two-wave nature of quark production, and suggest that the QGP was rather quark-rich. This latter conclusion could be better strengthened by better measurements of baryon yields, which differ from collaborations by several tens of percent. 

\begin{table}
\centerline{$f_B=0.5, n_{\rm quarks}/n_{\rm had}=0.85$}
\begin{tabular}{|r|c|c|c|c|c|c|c|c|c|c|c|} \hline
                  &           $p$ &     $\Lambda$ &    $\Sigma^+$ &    $\Sigma^-$ &       $\Xi^0$ &       $\Xi^-$ &    $\Omega^-$ &       $\pi^+$ &         $K^+$\\ \hline
       $\bar{p}$  &  0.535,-0.164 &  0.586,-0.260 &  0.591,-0.246 &  0.581,-0.273 &  0.643,-0.342 &  0.638,-0.355 &  0.694,-0.438 &  0.005, 0.013 & -0.052, 0.096\\ 
 $\bar{\Lambda}$  &  0.221,-0.098 &  0.278,-0.132 &  0.278,-0.132 &  0.278,-0.132 &  0.334,-0.167 &  0.334,-0.167 &  0.391,-0.201 & -0.000, 0.000 & -0.057, 0.034\\ 
$\bar{\Sigma}^-$  &  0.089,-0.037 &  0.111,-0.053 &  0.113,-0.049 &  0.110,-0.057 &  0.135,-0.065 &  0.133,-0.069 &  0.156,-0.081 &  0.002, 0.004 & -0.022, 0.016\\ 
$\bar{\Sigma}^+$  &  0.088,-0.041 &  0.111,-0.053 &  0.110,-0.057 &  0.113,-0.049 &  0.133,-0.069 &  0.135,-0.065 &  0.156,-0.081 & -0.002,-0.004 & -0.023, 0.012\\ 
   $\bar{\Xi}^0$  &  0.056,-0.030 &  0.077,-0.038 &  0.077,-0.037 &  0.076,-0.039 &  0.098,-0.046 &  0.098,-0.047 &  0.119,-0.054 &  0.000, 0.001 & -0.021, 0.009\\ 
   $\bar{\Xi}^+$  &  0.055,-0.031 &  0.077,-0.038 &  0.076,-0.039 &  0.077,-0.037 &  0.098,-0.047 &  0.098,-0.046 &  0.119,-0.054 & -0.000,-0.001 & -0.021, 0.008\\ 
$\bar{\Omega}^+$  &  0.011,-0.007 &  0.016,-0.008 &  0.016,-0.008 &  0.016,-0.008 &  0.021,-0.010 &  0.021,-0.010 &  0.027,-0.011 & -0.000,-0.000 & -0.005, 0.001\\ 
         $\pi^-$  &  0.121, 0.319 & -0.000, 0.000 &  0.242, 0.639 & -0.242,-0.639 &  0.121, 0.319 & -0.121,-0.319 & -0.000, 0.000 &  0.242, 0.639 &  0.121, 0.319\\ 
           $K^-$  & -0.248, 0.459 & -0.718, 0.437 & -0.694, 0.502 & -0.742, 0.373 & -1.164, 0.480 & -1.188, 0.415 & -1.634, 0.458 &  0.024, 0.064 &  0.470, 0.022\\ 
\hline
\end{tabular}
\centerline{$f_B=0.7, n_{\rm quarks}/n_{\rm had}=0.85$}
\begin{tabular}{|r|c|c|c|c|c|c|c|c|c|c|c|} \hline
                  &           $p$ &     $\Lambda$ &    $\Sigma^+$ &    $\Sigma^-$ &       $\Xi^0$ &       $\Xi^-$ &    $\Omega^-$ &       $\pi^+$ &         $K^+$\\ \hline
       $\bar{p}$  &  0.375, 0.004 &  0.411,-0.092 &  0.418,-0.074 &  0.405,-0.111 &  0.455,-0.170 &  0.448,-0.189 &  0.492,-0.266 &  0.007, 0.019 & -0.037, 0.096\\ 
 $\bar{\Lambda}$  &  0.155,-0.035 &  0.217,-0.067 &  0.217,-0.067 &  0.217,-0.067 &  0.279,-0.099 &  0.279,-0.099 &  0.341,-0.131 & -0.000, 0.000 & -0.062, 0.032\\ 
$\bar{\Sigma}^-$  &  0.063,-0.011 &  0.087,-0.027 &  0.089,-0.021 &  0.085,-0.032 &  0.113,-0.037 &  0.110,-0.042 &  0.136,-0.052 &  0.002, 0.006 & -0.024, 0.016\\ 
$\bar{\Sigma}^+$  &  0.061,-0.017 &  0.087,-0.027 &  0.085,-0.032 &  0.089,-0.021 &  0.110,-0.042 &  0.113,-0.037 &  0.136,-0.052 & -0.002,-0.006 & -0.026, 0.010\\ 
   $\bar{\Xi}^0$  &  0.039,-0.015 &  0.064,-0.023 &  0.065,-0.021 &  0.063,-0.024 &  0.089,-0.029 &  0.089,-0.031 &  0.114,-0.037 &  0.001, 0.002 & -0.025, 0.008\\ 
   $\bar{\Xi}^+$  &  0.039,-0.016 &  0.064,-0.023 &  0.063,-0.024 &  0.065,-0.021 &  0.089,-0.031 &  0.089,-0.029 &  0.114,-0.037 & -0.001,-0.002 & -0.025, 0.006\\ 
$\bar{\Omega}^+$  &  0.008,-0.004 &  0.014,-0.005 &  0.014,-0.005 &  0.014,-0.005 &  0.020,-0.007 &  0.020,-0.007 &  0.027,-0.008 & -0.000, 0.000 & -0.006, 0.001\\ 
         $\pi^-$  &  0.118, 0.317 & -0.000, 0.000 &  0.236, 0.634 & -0.236,-0.634 &  0.118, 0.317 & -0.118,-0.317 & -0.000, 0.000 &  0.236, 0.634 &  0.118, 0.317\\ 
           $K^-$  & -0.126, 0.331 & -0.560, 0.291 & -0.536, 0.355 & -0.584, 0.227 & -0.971, 0.315 & -0.994, 0.251 & -1.405, 0.275 &  0.024, 0.064 &  0.434, 0.040\\ 
\hline
\end{tabular}
\centerline{$f_B=0.6, n_{\rm quarks}/n_{\rm had}=0.7$}
\begin{tabular}{|r|c|c|c|c|c|c|c|c|c|c|c|} \hline
                  &           $p$ &     $\Lambda$ &    $\Sigma^+$ &    $\Sigma^-$ &       $\Xi^0$ &       $\Xi^-$ &    $\Omega^-$ &       $\pi^+$ &         $K^+$\\ \hline
       $\bar{p}$  &  0.363, 0.012 &  0.399,-0.076 &  0.404,-0.059 &  0.394,-0.093 &  0.440,-0.147 &  0.435,-0.165 &  0.476,-0.236 &  0.005, 0.017 & -0.036, 0.088\\ 
 $\bar{\Lambda}$  &  0.151,-0.029 &  0.200,-0.052 &  0.200,-0.052 &  0.200,-0.052 &  0.248,-0.075 &  0.248,-0.075 &  0.297,-0.097 &  0.000,-0.000 & -0.049, 0.023\\ 
$\bar{\Sigma}^-$  &  0.061,-0.009 &  0.080,-0.021 &  0.081,-0.016 &  0.078,-0.026 &  0.100,-0.027 &  0.099,-0.032 &  0.119,-0.039 &  0.001, 0.005 & -0.019, 0.012\\ 
$\bar{\Sigma}^+$  &  0.060,-0.014 &  0.080,-0.021 &  0.078,-0.026 &  0.081,-0.016 &  0.099,-0.032 &  0.100,-0.027 &  0.119,-0.039 & -0.001,-0.005 & -0.020, 0.007\\ 
   $\bar{\Xi}^0$  &  0.038,-0.013 &  0.057,-0.017 &  0.057,-0.016 &  0.057,-0.019 &  0.076,-0.020 &  0.076,-0.021 &  0.095,-0.024 &  0.000, 0.001 & -0.019, 0.004\\ 
   $\bar{\Xi}^+$  &  0.038,-0.014 &  0.057,-0.017 &  0.057,-0.019 &  0.057,-0.016 &  0.076,-0.021 &  0.076,-0.020 &  0.095,-0.024 & -0.000,-0.001 & -0.019, 0.003\\ 
$\bar{\Omega}^+$  &  0.007,-0.004 &  0.012,-0.004 &  0.012,-0.004 &  0.012,-0.004 &  0.017,-0.004 &  0.017,-0.004 &  0.022,-0.005 & -0.000,-0.000 & -0.005, 0.000\\ 
         $\pi^-$  &  0.098, 0.339 &  0.000,-0.000 &  0.196, 0.678 & -0.196,-0.678 &  0.098, 0.339 & -0.098,-0.339 & -0.000,-0.000 &  0.196, 0.678 &  0.098, 0.339\\ 
           $K^-$  & -0.144, 0.353 & -0.517, 0.242 & -0.497, 0.310 & -0.536, 0.173 & -0.869, 0.199 & -0.889, 0.131 & -1.241, 0.088 &  0.020, 0.068 &  0.372, 0.111\\ 
\hline
\end{tabular}
\centerline{$f_B=0.6, n_{\rm quarks}/n_{\rm had}=1.0$}
\begin{tabular}{|r|c|c|c|c|c|c|c|c|c|c|c|} \hline
                  &           $p$ &     $\Lambda$ &    $\Sigma^+$ &    $\Sigma^-$ &       $\Xi^0$ &       $\Xi^-$ &    $\Omega^-$ &       $\pi^+$ &         $K^+$\\ \hline
       $\bar{p}$  &  0.519,-0.144 &  0.570,-0.247 &  0.577,-0.232 &  0.563,-0.262 &  0.629,-0.336 &  0.622,-0.351 &  0.681,-0.440 &  0.007, 0.015 & -0.052, 0.104\\ 
 $\bar{\Lambda}$  &  0.215,-0.093 &  0.285,-0.137 &  0.285,-0.137 &  0.285,-0.137 &  0.355,-0.181 &  0.355,-0.181 &  0.425,-0.225 &  0.000,-0.000 & -0.070, 0.044\\ 
$\bar{\Sigma}^-$  &  0.087,-0.035 &  0.114,-0.055 &  0.116,-0.050 &  0.112,-0.059 &  0.143,-0.070 &  0.141,-0.075 &  0.170,-0.090 &  0.002, 0.005 & -0.027, 0.020\\ 
$\bar{\Sigma}^+$  &  0.085,-0.040 &  0.114,-0.055 &  0.112,-0.059 &  0.116,-0.050 &  0.141,-0.075 &  0.143,-0.070 &  0.170,-0.090 & -0.002,-0.005 & -0.029, 0.015\\ 
   $\bar{\Xi}^0$  &  0.055,-0.029 &  0.081,-0.042 &  0.082,-0.040 &  0.081,-0.043 &  0.109,-0.053 &  0.108,-0.054 &  0.136,-0.065 &  0.001, 0.001 & -0.027, 0.012\\ 
   $\bar{\Xi}^+$  &  0.054,-0.030 &  0.081,-0.042 &  0.081,-0.043 &  0.082,-0.040 &  0.108,-0.054 &  0.109,-0.053 &  0.136,-0.065 & -0.001,-0.001 & -0.028, 0.011\\ 
$\bar{\Omega}^+$  &  0.011,-0.007 &  0.017,-0.009 &  0.017,-0.009 &  0.017,-0.009 &  0.024,-0.012 &  0.024,-0.012 &  0.031,-0.014 & -0.000,-0.000 & -0.007, 0.002\\ 
         $\pi^-$  &  0.140, 0.297 &  0.000,-0.000 &  0.281, 0.594 & -0.281,-0.594 &  0.140, 0.297 & -0.140,-0.297 &  0.000,-0.000 &  0.281, 0.594 &  0.140, 0.297\\ 
           $K^-$  & -0.206, 0.415 & -0.738, 0.463 & -0.710, 0.523 & -0.766, 0.403 & -1.241, 0.571 & -1.269, 0.512 & -1.773, 0.620 &  0.028, 0.060 &  0.532,-0.048\\ 
\hline
\end{tabular}
\caption{\label{table:bfudgenq}
Weights $w_{\alpha\beta}^{\rm(QGP)}$ and $w_{\alpha\beta}^{\rm(had)}$ as in Table \ref{table:default} except for a higher and lower yields of baryon number (adjusted by the scaling factor $f_B$), and quark densities before hadronization, $n_{\rm quarks}$. The upper two tables show weights for variations of $f_B$ from the default value of 0.6 to 0.5 and 0.7, while the lower two tables display results for varying $n_{\rm quarks}/n_{\rm had}$ from the default value of 0.85 to 0.7 and 1.0. The $p\bar{p}$ weights are considerably sensitive to both the baryon yield and the input baryon density. Either higher final-state baryon yields, or lower quark densities strengthen the hadronization peak in the $p\bar{p}$ balance function.  }
\end{table}

\section{Blast-wave predictions for balance functions}

Once one has calculated the weights described in the previous section, one can calculate the balance function between any two species in coordinate space given the characteristic widths of the distributions, $\sigma_{\rm(QGP)}$ and $\sigma_{\rm(had)}$. There is no firm understanding of the scale $\sigma_{\rm(QGP)}$, as the value depends on the microscopic details of how quark-antiquark pairs are created in the pre-thermalization stage. In a flux-tube picture, the quarks are pulled apart longitudinally, using the tube's energy to create the particles. From balance functions of $pp$ collisions, one would estimate $\sigma_{\rm(QGP)}\gtrsim 0.5$ units of rapidity. Even if the balancing quark pairs are created atop one another, one would still expect the range to be near 0.5 once one accounted for diffusion, which spreads logarithmically with the time \cite{Bass:2000az}. The characteristic spread for $\sigma_{\rm(had)}$ should be determined by the diffusion that occurs after hadronization. Although the time from hadronization ($\sim 7$ fm/$c$) to breakup ($\sim 14$ fm/$c$) is similar as the time from creation to hadronization, the  diffusion width grows logarithmically with time, and the post hadronization diffusion should be $\lesssim 0.2$ units of rapidity. Since the spread from final-state thermal motion is likely larger than $\sigma_{\rm(had)}$, the choice of 0.1 vs 0.2 for the width should not strongly impact the results. Since the purpose of this section is to provide an example providing a crude idea of what one might suspect, the values are picked with some arbitrariness to be $\sigma_{\rm(QGP)}=0.6,~\sigma_{\rm(had)}=0.2$, i.e.,
\begin{equation}
B_{\alpha,\beta}(\Delta\eta)=\frac{w_{\alpha\beta}^{\rm(had)}}{(2\pi)^{1/2}\sigma_{\rm(had)}}e^{-(\Delta\eta)^2/(2\sigma_{\rm(had)}^2)}
+\frac{w_{\alpha\beta}^{\rm(QGP)}}{(2\pi)^{1/2}\sigma_{\rm(QGP)}}e^{-(\Delta\eta)^2/(2\sigma_{\rm(QGP)}^2)}.
\end{equation}

Unfortunately, the balance function is not measured in coordinate space. The mapping of $\eta\rightarrow y$ has a spread from the thermal motion of the particles at breakup. For pions this can be a half unit of rapidity, whereas for protons the thermal spread is only a few tenths. Additionally, particles decay. To include both decays and the thermal spread, the correlations in coordinate space were overlaid onto a simple blast-wave parameterization. The blast-wave parameterization models the collective and thermal motion by assuming that the radial flow grows linearly in radius, $u_i=u_{\rm max}r_i/r_{\rm max}$, with $u_{\rm max}=1.0$, and that the breakup temperature is 100 MeV. Decays of unstable particles (lambdas, neutral kaons, Sigmas, Cascades and Omegas) were also accounted for by a Monte-Carlo simulation. 

\begin{figure}
\centerline{\includegraphics[width=0.5\textwidth]{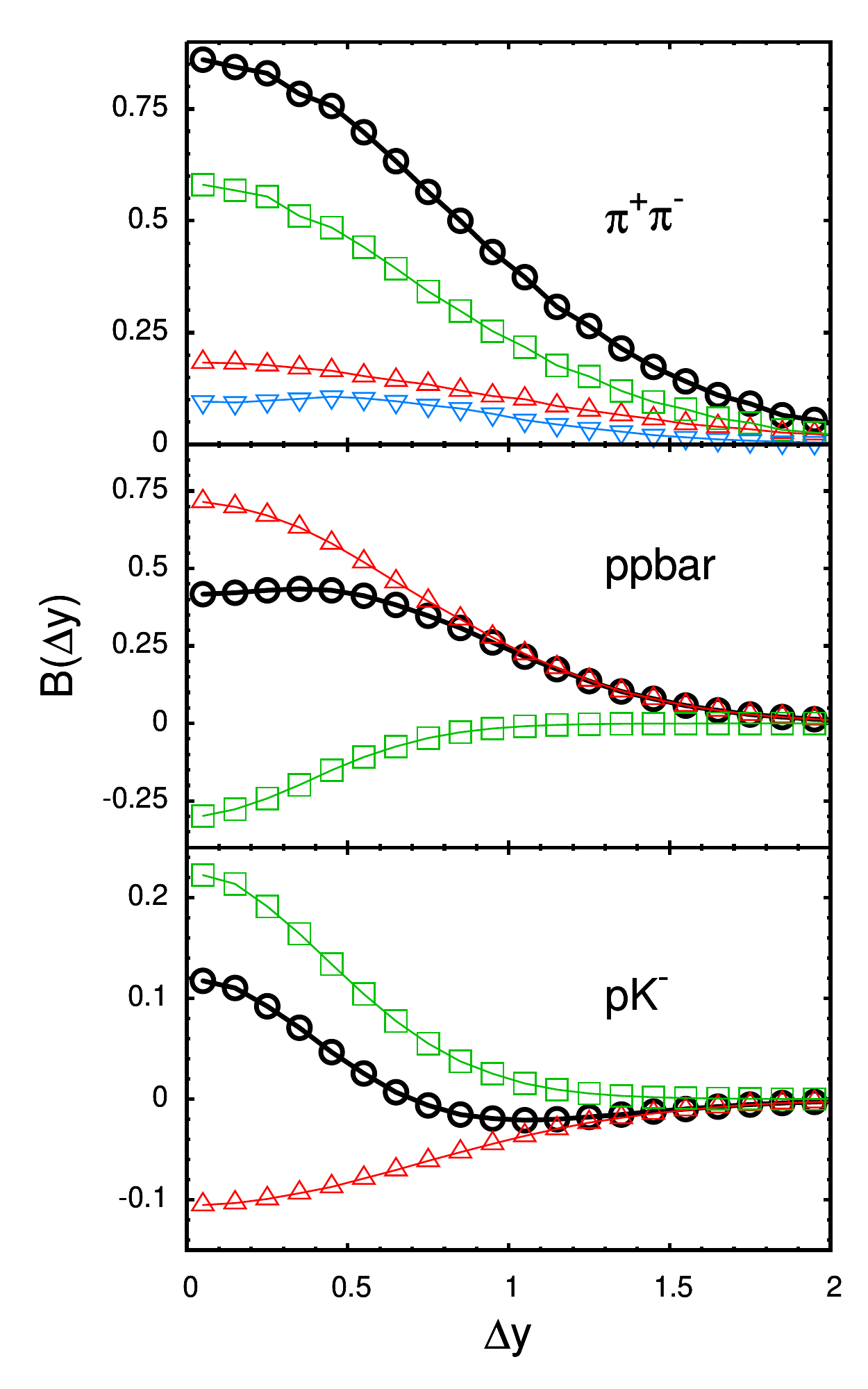}}
\caption{\label{fig:blastwave}(color online)
Balance functions for $\pi^+\pi^-$ are shown as a function of relative rapidity (black circles) in the upper panel as calculated with the default values given in Table \ref{table:default}. The widths assumed are $\sigma_{\rm(had)}=0.2,~\sigma_{\rm(QGP)}=0.6$. The blast-wave model is used to map $B(\Delta\eta)$ to $B(\Delta y)$. The hadronization component (green squares) is larger and narrower than the QGP component (red upward triangles), due to the fact that most quark-antiquark pairs are created at hadronization. The contribution from weak decays (blue downward triangles) to two pions is also shown. Since the QGP contribution is rather small and since the thermal spread for pions is large, one can not distinguish the two components, but must be content with the width being consistent with being dominated by a single narrow contribution. In contrast, the $p\bar{p}$ and $pK^-$ balance functions shown in the middle and lower panels clearly illustrate the two-component nature. Because the $p\bar{p}$ balance function has contributions with opposite signs, one can expect to see a plateau or even a dip at small relative $\Delta y$. Since the $p\bar{p}$ balance function is dominated by the QGP contribution while the $\pi^+\pi^-$ balance function is largely driven by the hadronization contribution, one would expect the apparent width $\sigma_\eta$ for a single-wave fit to be much broader for the $p\bar{p}$ case than for the $\pi^+\pi^-$ case. For the $pK^-$ balance functions, adding the two contributions results in a balance function that is narrower than what one could get from a model with a single wave. Further, one may even see the balance function become negative for $\Delta y\sim 1.0$.
}
\end{figure}

Three of the resulting balance functions are presented in Fig. \ref{fig:blastwave}, and are broken down by components. The $\pi^+\pi^-$ balance function is dominated by the hadronization component, with the QGP component contributing to the tail. The contribution from final-state decays is small, but non-negligible. Due to large thermal spread for pions, it is difficult to distinguish the two components. 

The $p\bar{p}$ balance function, displayed in the middle panel of Fig. \ref{fig:blastwave}, is dominated by the QGP component, with the hadronization component being small or negative. The calculation includes the contribution to protons from weak decays, and the hadronization component from hyperon balance functions makes the hadronization component more negative. Higher quark densities in the QGP or lower final-state baryon yields push this component toward being negative. If it is negative, as in the case of the default calculation in Fig. \ref{fig:blastwave}, the resulting balance function has a plateau or perhaps a dip at small $\Delta y$. The existence of such a dip would provide striking evidence for the two-wave nature of charge production. If the hadronization component were zero, one could still see evidence of two components by comparing with the $\pi^+\pi^-$ balance functions. Since the $p\bar{p}$ balance function is dominated by the QGP component while $\pi^+\pi^-$ balance function is driven by the hadronization component, one could perform a single-wave fit to the width of the balance function in coordinate space $\sigma_\eta$. One would expect the width for $p\bar{p}$ to be significantly larger than that for $\pi^+\pi^-$. The width, $\sigma_\eta$ for $\pi^+\pi^-$ in coordinate space has been determined by a blast-wave analysis in   \cite{Schlichting:2010qia}. By using blast-wave parameters fit to spectra and elliptic flow observables, the analysis determined that the width of the balance function in coordinate space, assuming a single scale for the charge correlation, was $\sigma_\eta\sim 0.22$ for the most central collisions. By performing the blast-wave fit, the contribution from final-state thermal motion was effectively subtracted to find the width in coordinate space. This width fell for increasing centrality from 0.6 to nearly 0.2. For the $p\bar{p}$ balance function one might see $\sigma_\eta$ stay roughly constant with centrality.

The $pK^-$ balance function, from the lower panel of Fig. (\ref{fig:blastwave}), offers yet more promise for demonstrating the two-component nature of the balance function. Since the weights of the two components have different signs with the stronger component having the smaller width, one finds both positive and negative regions of the balance function. If there were only one component, or if the two components had similar widths, this behavior would not ensue. Further, since protons and kaons are more massive, the thermal spread is reduced and the reduction in smearing allows more resolving power into the correlations in coordinate space. The negative dip for $\Delta y\sim 1.0$ might be reduced if the two waves of charge production are not well separated. For instance, if hadronization was more gradual the narrow peak would have a long non-Gaussian tail which could overwhelm the smaller negative contribution from the QGP component. However, even in this case the overall width of the $pK^-$ balance function would be narrowed by the second component. Thus, the signal for two waves of charge production which are only semi-distinct would be a narrow $pK^-$ balance function, whose width might even be narrower that what one would predict from a single-wave model with zero width after one corrects for thermal broadening. The difficulty with $pK^-$ balance functions comes from the fact that they are smaller, by nearly an order of magnitude, than the $\pi^+\pi^-$ balance functions, and thus require high-statistics data sets. Fortunately, both STAR at RHIC and ALICE at the LHC provide both high statistics, and due to the installation of large-coverage time-of-flight detectors, large acceptances for identified particles.

\section{Summary}

A central feature of the canonical picture of the chemical evolution of the quark-gluon plasma is the two-wave nature of quark production. Investigating balance functions over a large range of species pairs provide the means to test this hypothesis in great detail. Once baryon production is better understood, the only parameter affecting the calculation of weights is the quark density in the QGP. The blast-wave parameters used to model the thermal broadening of the balance-function structures are already well determined by spectra. This leaves three parameters, $n_{\rm quarks}/n_{\rm had}$, $\sigma_{\rm(had)}$ and $\sigma_{\rm(QGP)}$ for fitting the entire array of balance functions. If one were to also question the assumption that the strange quark density was close ($\simeq 90$\%) to the up or down quark density, one would add a fourth parameter.

The existence of two waves of charge production could have several clear signatures.
\begin{itemize}\itemsep 0 pt
\item The width of the $\pi^+\pi^-$ balance function in $\Delta\eta$ (coordinate space rapidity) should be small for central collisions. This has been reported in \cite{Schlichting:2010qia}.
\item In central collisions the width of the $p\bar{p}$ and $K^+K^-$ balance functions in $\Delta\eta$ should be larger than that of the $\pi^+\pi^-$ balance function. Whereas the width for pions has been observed to shrink with centrality, these widths may well stay fixed, or even broaden for increasing centrality.
\item The $p\bar{p}$ balance function could have a plateau or even a dip at small $\Delta y$. 
\item The $pK^-$ balance function should be narrower than can be fit with a single-wave picture, and might dip negative for $\Delta y\sim 1.0$.
\end{itemize}

Aside from qualitatively demonstrating the two-wave nature of quark production, the numerical parameters one might extract by fitting to data are also of high interest. To date, there has not been a convincing means for extracting the number density of quarks in the plasma, $n_{\rm quarks}$, from experiment. Determining the width, $\sigma_{\rm(QGP)}$, would provide insight into the dynamical mechanism for the creation and diffusion of quarks in the plasma.

Potentially, the most important implication of charge balance function would be to quantitatively constrain the charge correlations in the QGP. For this study, the density of quarks was varied, which then determined the magnitude of the diagonal components of $g_{ab}(\Delta\eta)$ in the QGP. Several of the hadronic balance functions were then found to be sensitive to this number. Additionally, there was an explicit assumption that the off-diagonal elements were zero in the QGP. This would not be the case if quark-antiquark pairs, such as pionic fluctuations, made significant contributions to the entropy of the QGP. Observing that the off-diagonal elements were small or zero, would make a strong case that quarks are the dominant quasi-particles in the QGP. In principle, one could extend the ideas presented here and vary the off-diagonal elements to determine the ranges to which they are constrained by experiment.

The calculations presented here are somewhat schematic in nature, and can be improved during the coming years. Most immediately, the list of resonances considered was small, and omitted the short-lived hadronic states such as the $\rho$ or $\Delta$, which should have some measurable effect \cite{Florkowski:2004em}. Although almost all particles have a decay in their history, on the order of 10\% of the charged particles produced at RHIC come from the decays of neutral resonances, other than the weak decays accounted for here, where both charges escape untouched from the decay. Thus, several of the weights might be affected at the 10\% level in a more thorough calculation. If the data indeed seems addressable with this schematic model, one could consider more sophisticated models of quark production, diffusion, hadronization and emission.

\end{document}